********************************************************************
This starts the beginning of the paper.  No special macros are needed.
*********************************************************************

\documentstyle[12pt]{article}
\newcommand{\beq}{\begin{equation}}
\newcommand{\eeq}{\end{equation}}
\newcommand{\beqa}{\begin{eqnarray}}
\newcommand{\eeqa}{\end{eqnarray}}
\newcommand{\labeq}[1]{\label{eq:#1}}
\newcommand{\refeq}[1]{\ref{eq:#1}}
\newcommand{\labref}[1]{\label{ref:#1}}
\newcommand{\rref}[1]{\ref{ref:#1}}

\newcommand{\bdm}{\begin{displaymath}}
\newcommand{\edm}{\end{displaymath}}
\newcommand{\bdma}{\begin{eqnarray*}}
\newcommand{\edma}{\end{eqnarray*}}

\begin{document}

\begin{flushright}
DOE-287-CPP-25
\end{flushright}

\vspace*{24pt}

\begin{center}
{\large{\bf Toward a Conformal Field Theory\\ for the Quantum Hall Effect}}\\
\vspace{24pt}
Greg Nagao\\
Center for Particle Physics\\
The University of Texas\\
Austin, Texas~~78712
\end{center}
\vspace{64pt}

\centerline{\bf Abstract}
\vspace{12pt}

An effective Hamiltonian for the study of the quantum Hall effect is
proposed.  This Hamiltonian, which includes a ``current-current" interaction
has the form of a Hamiltonian for a conformal field theory in the large $N$
limit.  An order parameter is constructed from which the Hamiltonian may
be derived.  This order parameter may be viewed as either a collective
coordinate for a system of $N$ charged particles in a strong magnetic field;
or as a field of spins associated with the cyclotron motion of these particles.

\newpage
\noindent
{\bf I.~~Overview}
\vspace{12pt}

The IQHE is described by the Hamiltonian for a single particle in an external
magnetic field.  On the other hand, the FQHE is considered as a manifestation
of a many-body effect of $N$ interacting particles in an external magnetic
field [\rref{L3}].  The center of mass motion of these $N$ particles are
usually neglected, assuming the fluctuations about the center of mass contain
the key features neccessary to understand the fractional quantization of the
Hall conductivity.

There have been many recent papers (e.g. [\rref{F2},\rref{S1}])
relating vertex operators to the quantum Hall effect.  In particular, it has
been shown that the Laughlin wavefunction may be obtained as an $N'$-
point correlation function of vertex operators.  The conformal weights of
these vertex operators are associated with fractional charges of the
quasiparticle states.  An attempt is made here to better understand these
results in the context of a 2+1 dimensional conformal field theory (CFT).
Recalling  that the conformal weights of vertex operators in string theory are
associated with the center of mass (COM) motion of the string, we explicitly
study the separation of the COM motion for a system of $N$ charged particles
in a strong magnetic field.   We find that the $N$-particle Hamiltonian
(including the COM motion) may be diagonalized through the inclusion of a
``current-current" interaction.  It is then shown that this leads  to an
effective Hamiltonian resembling that of a CFT in the large $N$-limit.  This
suggests that we define a collective coordinate $\varphi_L (\zeta)$ in
symmetric gauge.  This collective coordinate may be viewed as a field of
spins associated with the cyclotron motion of charged particles in the strong
magnetic field.  This ``field of spins" may be taken as an
order parameter for the study of the vacuum structure of the QHE.

\vspace{24pt}
\noindent
{\bf II.~~$N$ Non-interacting Particles in an External Magnetic Field}
\vspace{12pt}

The Hamiltonian usually taken to describe the QHE is that of $N$
noninteracting particles in an external magnetic field

\beq
H_0 =  \frac{1}{2\mu} \sum_{n=1}^N \vec{\pi}_n^2
\eeq

\noindent
where
\beqa
\vec{\pi}_n  \equiv & \vec{p}_n + \frac{e}{c} \vec{A}(\vec{x}_n)\\
A^i(\vec{x}_n) \equiv & \frac{B}{2} \left(\epsilon^{ij} x^j_n + \frac{\partial
\Lambda}{\partial x^i_n} \right)
\eeqa

\noindent
are the canonical momentum and vector potential in arbitrary
gauge.  Separating out the center of mass energy, the Hamiltonian becomes

\beq
H_0 =  \frac{1}{2\mu} \left( \vec{\pi}_0^2 - \sum_{n\neq m = 1}^N \vec{\pi}_n
\cdot \vec{\pi}_m \right)
\eeq

\noindent
where the center of mass (COM) coordinates are

\beqa
\vec{p}_0 \equiv \sum_{n=1}^N \vec{p}_n~~~~~&;&~~~~~ \vec{x}_0 \equiv
\frac{1}{N} \sum_{n=1}^N \vec{x}_n\\
\vec{\pi}_0 \equiv \vec{p}_0 + \frac{e}{c}
\vec{A}(N\vec{x}_0) &\Longrightarrow&
\pi^i_0 = p^i_0  + \frac{\hbar N}{2\ell^2} \left( \epsilon^{ij} x^j_0 +
\frac{1}{N^2} \frac{\partial \Lambda}{\partial x^i_0} \right).
\eeqa

\noindent
Performing a canonical transformation to complex phase space coordinates,
the Hamiltonian may be written as

\beq
H_0 = \hbar \omega \left( N \alpha^*_0 \alpha_0 - \sum_{n\neq m = 1}^N
\alpha^*_n \alpha_m \right)
\eeq

\noindent
where $\omega\equiv \frac{eB}{\mu c}$ is the cyclotron frequency, $\ell^2
\equiv \frac{eB}{\mu c}$ is the magnetic length and
$\alpha$ is a dimensionless complex momentum

\beqa
\alpha_0 \equiv \frac{\ell}{\sqrt{2N} \hbar} \left( \pi_0^1 + i \pi_0^2
\right)~~~~~&;&~~~~~
\alpha_0^* \equiv \frac{\ell}{\sqrt{2N} \hbar} \left( \pi_0^1 - i \pi_0^2
\right)\\
\alpha_n \equiv \frac{\ell}{\sqrt{2} \hbar} \left( \pi_n^1 + i \pi_n^2
\right)~~~~~&;&~~~~~
\alpha_n^* \equiv \frac{\ell}{\sqrt{2} \hbar} \left( \pi_n^1 - i \pi_n^2
\right)\nonumber
\eeqa

\noindent
with commutation relations (using $\{x_0^i,p_0^j\} = \delta^{ij}$ and
$\{x_n^i,p_n^j\} = \delta^{ij}$ )

\beqa
\{\pi_0^i,\pi_0^j\} &=& \frac{N\hbar}{2\ell^2} \left[ 2 \epsilon^{ij} +
\frac{1}{N^2} \left( \left\{p_0^i,
\frac{\partial \Lambda}{\partial x_0^j}\right\} + \left\{\frac{\partial
\Lambda} {\partial x_0^j} , p_0^j\right\} \right) \right]\nonumber\\
\{\pi_n^i,\pi_n^j\} &=& \frac{\hbar}{2\ell^2} \left[ 2 \epsilon^{ij} +
\left\{p_n^i,
\frac{\partial \Lambda}{\partial x_n^j}\right\} + \left\{\frac{\partial
\Lambda} {\partial x_n^j} , p_n^j\right\} \right]\nonumber\\
\{\alpha_0 , \alpha_0^*\} &=& \frac{-i}{2\hbar} \left( 2 \{x_0^1,p_0^1\} +
\frac{1}{N^2}
\left\{\frac{\partial \Lambda}{\partial x_0^1} , p_0^2 \right\} - \frac{1}{N^2}
\left\{\frac{\partial \Lambda}{\partial x_0^2} , p_0^1 \right\}
\right)\nonumber\\
\{\alpha_n , \alpha_n^*\} &=& \frac{-i}{2\hbar} \left( 2 \{x_n^1,p_n^1\} +
\left\{\frac{\partial
\Lambda}{\partial x_n^1} , p_n^2 \right\} -  \left\{\frac{\partial
\Lambda}{\partial x_n^2} , p_n^1 \right\} \right).\nonumber
\eeqa

\noindent
In symmetric gauge $(\Lambda = 0)$ we thus have the expected
commutator results ($[x,p] = i\hbar$) for annihilation-creation operators:

\beqa
\{\pi_0^i, \pi_0^j\} = \frac{N\hbar}{\ell^2} \epsilon^{ij}~~~&;&~~~
\{\pi_n^i, \pi_n^j\} \frac{\hbar}{\ell^2} \epsilon^{ij}\nonumber\\
\left[\pi_0^i,\pi_0^j\right] = iN \left(\frac{\hbar}{\ell}\right)^2
\epsilon^{ij} ~~~&;&~~~
\left[\pi_n^i,\pi_n^j\right] = i \left(\frac{\hbar}{\ell}\right)^2
\epsilon^{ij}\\
\{\alpha_0, \alpha_0^*\} = \frac{-i}{\hbar}~~~&;&~~~\{\alpha_n, \alpha_n^*\}
= \frac{-i}{\hbar}\nonumber\\
\left[\alpha_0 , \alpha_0^\dagger\right] = \frac{-i}{\hbar}
\left[x_0^1 , p_0^1\right] = 1 ~~~&;&~~~ \left[\alpha_n ,
\alpha_n^\dagger\right] = \frac{-i}{\hbar} \left[x_n^1 , p_n^1\right] =
1.\nonumber
\eeqa

\vspace{24pt}
\noindent
{\bf III.~~A Current-Current Interaction}
\vspace{12pt}

We now consider what happens if we include an interaction of the form

\beq
\frac{1}{2\mu} \sum_{n,m=1}^N \vec{\pi}_n \cdot \vec{\pi}_m \labeq{interaction}
\eeq

\noindent
will thus have the effect of diagonalizing the Hamiltonian

\beqa
H &=& H_0 + \frac{1}{2\mu} \sum_{n,m=1}^N \vec{\pi}_n \cdot
\vec{\pi}_m\\
 &=& \frac{1}{2\mu} \left[ \vec{\pi}_0^2 - \sum_{n\neq m=1}^N \vec{\pi}_n
\cdot \vec{\pi}_m + \sum_{n,m=1}^N \vec{\pi}_n \cdot \vec{\pi}_m \right]\\
 &=& \frac{1}{2\mu} \left( \vec{\pi}_0^2 + \sum_{n=1}^N \vec{\pi}_n^2
\right)
\eeqa

\noindent
This Hamiltonian can then be recognized as the generator of dilations ($L_0$)
for some Virasoro algebra (in the large $N$ limit) with $N$ playing the role
of the inverse string tension ($\alpha' \sim \frac{1}{T}$).  This can be most
easily seen in terms of our complex momentum $\alpha$ :

\beq
H = \hbar \omega \left( N \alpha_0^* \alpha_0 + \sum_{n=1}^N \alpha_n^*
\alpha_n \right).
\eeq

In terms of an effective field theory (i.e. in the large $N$-limit), such a
momentum-momentum coupling (\refeq{interaction}) becomes a current-current
interaction

\beq
\vec{\pi}\cdot \vec{\pi} \longrightarrow \vec{j} \cdot \vec{j}.
\eeq

\noindent
As we shall see (\refeq{current}), such a current may be related to the
conjugate momentum of a field associated with the collective coordinate for
the $N$-particle system.  We may thus identify this interaction as that which
is neccessary to describe this system in terms of a CFT.  The current here may
be recognized as that of a Kac-Moody algebra\footnote{The involvement of a
$U(1)$ Kac-Moody algebra in the QHE was first
recognized by Wen [\rref{Wen1}] in a different manner.} whose stress-energy
tensor is given by the Sugawara construction.  Since the only charges which
are present in this model is that of the electric charge (which enters into
the bosonic annihilation-creation operators),\footnote{The fermionic character
of the electrons has been neglected in this quantization.  It may be taken
into account through statistical methods when calculating such physical
quantities as the conductivity tensor.} we expect this current to be the
generator of a $U(1)$ Kac-Moody algebra.

\vspace{24pt}
\noindent
{\bf IV.~~A Quantum Representation in Complex Phase Space}
\vspace{12pt}

Our effective Hamiltonian of the previous section contains the energy
associated with COM momenta as well as that for fluctuations off the COM.
Until now we have not found it neccessary to introduce
configuration space variables.  It has been enough to define commutation
relations between conjugate momenta (or complex conjugate momenta).  In
order to give an explicit realization of this commutator algebra, however, it
is
convenient to introduce configuration space variables.  In particular, since it
is suggested that conformal symmetry plays an important role in this
system, we find it convenient to define complex configuration space
variables.  Note configuration space variables enter directly into the
Hamiltonian only through the vector potential.  Defining a complex vector
potential

\beq
A\equiv A^1 + i A^2
\eeq

\noindent
with

\beq
A^i = \frac{B}{2} \left( \epsilon^{ij} x^j + \frac{\partial \Lambda}{\partial
x^i}
\right).
\eeq

\noindent
allows us to evaluate the system in arbitrary gauge in terms of complex
variables

\beqa
\frac{e}{c} A &=& \frac{\hbar}{\ell} \left( -iz + \frac{1}{2\ell^2}
\frac{\partial \Lambda}{\partial z^*} \right)\\
z &\equiv & \frac{1}{2\ell} \left( x^1 + i x^2 \right).\nonumber
\eeqa

\noindent
This complex coordinate can thus be thought of as representing the position
of a magnetic flux quantum.  This can be seen more explicitly by noting that
classically, the magnetic flux quantum is located at the bariocentric
coordinate (or what we shall call a ``psuedohole" associated with the charge
conjugate operator [\rref{N1}]) around which the charge circulates.  The
bariocentric coordinate is given by

\beq
\vec{\tilde{\pi}} \equiv \vec{p} - \frac{e}{c} \vec{A} (\vec{x}) = {\cal C}
\vec{\pi}
\eeq

\noindent
where ${\cal C}$ is the charge conjugation operator; or in complex
momentum coordinates

\beqa
\beta &\equiv& \frac{\ell}{\sqrt{2} \hbar} \left( \tilde{\pi}^2 + i
\tilde{\pi}^1
\right)\\
 &=& i \alpha^* + \sqrt{2} \left( z^* - \frac{i}{2\ell^2} \frac{\partial
\Lambda}{\partial z} \right).\nonumber
\eeqa

\noindent
We may thus define coordinates for the COM and fluctuations off the COM
coordinates

\beqa
z_0 \equiv  \frac{N}{2\ell} \left( x_0^1 + i x_0^2 \right) ~~~~~&;&~~~~~
z_0^* = \frac{N}{2\ell} \left( x_0^1 - i x_0^2 \right)\nonumber\\
z_n \equiv \frac{1}{2\ell} \left( x_n^1 + i x_n^2 \right) ~~~~~&;&~~~~~
z_n^* = \frac{1}{2\ell} \left( x_n^1 - i x_n^2 \right)\nonumber\\
\rho_0 \equiv  \frac{\ell}{N} \left( p_0^1 - i p_0^2 \right) ~~~~~&;&~~~~~
\rho_0^* = \frac{\ell}{N} \left( p_0^1 + i p_0^2 \right)\\
\rho_n \equiv  \ell \left( p_n^1 - i p_n^2 \right) ~~~~~&;&~~~~~
\rho_n^* = \ell \left( p_n^1 + i p_n^2 \right)\nonumber\\
p_0^1 = \frac{N}{2\ell} \left( \rho_0 + \rho_0^* \right) ~~~~~&;&~~~~~
p_0^2 = \frac{iN}{2\ell} \left( \rho_0 - \rho_0^* \right)\nonumber
\eeqa

For the ``quasiparticle" in symmetric gauge, we then have

\beqa
\pi_0^1 = \frac{N}{2\ell} \left[ (\rho_0 + \rho_0^*) + \frac{i\hbar}{N}
(z_0^* - z_0) \right]~~~&;&~~~
\pi_0^2 = \frac{iN}{2\ell} \left[ (\rho_0 - \rho_0^*) + \frac{i\hbar}{N}
(z_0^* + z_0) \right]\nonumber\\
\pi_n^1 = \frac{1}{2\ell} \left[ (\rho_n + \rho_n^*) + i\hbar (z_n^* - z_n)
\right]~~~&;&~~~
\pi_n^2 = \frac{i}{2\ell} \left[ (\rho_n - \rho_n^*) + i\hbar (z_n^* + z_n)
\right]\\
\alpha_0 = \sqrt{\frac{N}{2}} \frac{1}{\hbar} \left( \rho_0^* -
\frac{i\hbar}{N} z_0 \right)~~~&;&~~~
\alpha_0^* = \sqrt{\frac{N}{2}} \frac{1}{\hbar} \left( \rho_0 +
\frac{i\hbar}{N} z_0^* \right)\nonumber\\
\alpha_n = \frac{1}{\sqrt{2}\hbar} \left( \rho_n^* - i\hbar z_n
\right)~~~&;&~~~
\alpha_n^* = \frac{1}{\sqrt{2}\hbar} \left( \rho_n + i\hbar z_n^*
\right);\nonumber
\eeqa

\noindent
and for the ``quasipsuedohole"

\beqa
\tilde{\pi}_0^1 = \frac{N}{2\ell} \left[ (\rho_0 + \rho_0^*) - \frac{i\hbar}{N}
(z_0^* - z_0) \right]~~~&;&~~~
\tilde{\pi}_0^2 = \frac{iN}{2\ell} \left[ (\rho_0 - \rho_0^*) -
\frac{i\hbar}{N}
(z_0^* + z_0) \right]\nonumber\\
\tilde{\pi}_n^1 = \frac{1}{2\ell} \left[ (\rho_n + \rho_n^*) - i\hbar (z_n^* -
z_n) \right]~~~&;&~~~
\tilde{\pi}_n^2 = \frac{i}{2\ell} \left[ (\rho_n - \rho_n^*) - i\hbar (z_n^* +
z_n) \right]\\
\beta_0 = \sqrt{\frac{N}{2}} \frac{1}{\hbar} \left( i\rho_0 +
\frac{\hbar}{N} z_0^* \right)~~~&;&~~~
\beta_0^* = \sqrt{\frac{N}{2}} \frac{1}{\hbar} \left( -i\rho_0^* +
\frac{\hbar}{N} z_0 \right)\nonumber\\
\beta_n = \frac{1}{\sqrt{2}\hbar} \left( i\rho_n + \hbar z_n^*
\right)~~~&;&~~~
\beta_n^* = \frac{1}{\sqrt{2}\hbar} \left( -i\rho_n^* + \hbar z_n
\right).\nonumber
\eeqa

Differential operators may be defined

\beqa
\frac{\partial}{\partial z_0} \equiv \frac{\ell}{N} \left(
\frac{\partial}{\partial x_0^1} - i \frac{\partial}{\partial x_0^2}
\right)~~~~~&;&~~~~~\frac{\partial}{\partial z_0^*} = \frac{\ell}{N} \left(
\frac{\partial}{\partial x_0^1} + i \frac{\partial}{\partial x_0^2} \right)\\
\frac{\partial}{\partial z_n} \equiv \ell \left( \frac{\partial}{\partial
x_n^1} - i \frac{\partial}{\partial x_n^2}
\right)~~~~~&;&~~~~~\frac{\partial}{\partial z_n^*} = \ell \left(
\frac{\partial}{\partial x_n^1} + i \frac{\partial}{\partial x_n^2}
\right)\nonumber
\eeqa

\noindent
giving rise to the usual quantum realizations

\beqa
\rho_0 = -i\hbar \frac{\partial}{\partial z_0}~~~~~&;&~~~~~\rho_0^* = -
i\hbar \frac{\partial}{\partial z_0^*}\\
\rho_n = -i\hbar \frac{\partial}{\partial z_n}~~~~~&;&~~~~~\rho_n^* = -
i\hbar \frac{\partial}{\partial z_n^*}\nonumber
\eeqa

\noindent
and quantum commutators

\beqa
\left[z_0,\rho_0\right] = i\hbar~~~~~&;&~~~~~\left[z_0^*,\rho_0^*\right]
 = i\hbar\\
\left[z_n,\rho_n\right] = i\hbar~~~~~&;&~~~~~\left[z_n^*,\rho_n^*\right] =
i\hbar\nonumber
\eeqa

\noindent
with all others vanishing.  Our complex ``momentum" operators become

\beqa
\alpha_0 = -i\sqrt{\frac{N}{2}} \left( \frac{\partial}{\partial z_0^*} +
\frac{1}{N} z_0 \right)~~~~~&;&~~~~~
\alpha_0^\dagger = -i\sqrt{\frac{N}{2}} \left( \frac{\partial}{\partial z_0}
- \frac{1}{N} z_0^* \right)\nonumber\\
\alpha_n = \frac{-i}{\sqrt{2}} \left( \frac{\partial}{\partial z_n^*} +
 z_n \right)~~~~~&;&~~~~~
\alpha_n^\dagger = \frac{-i}{\sqrt{2}} \left( \frac{\partial}{\partial z_n} -
 z_n^* \right)\\
\beta_0 = -\sqrt{\frac{N}{2}} \left( \frac{\partial}{\partial z_0} +
\frac{1}{N} z_0^* \right)~~~~~&;&~~~~~
\beta_0^\dagger = -\sqrt{\frac{N}{2}} \left( \frac{\partial}{\partial z_0^*}
- \frac{1}{N} z_0 \right)\nonumber\\
\beta_n = \frac{1}{\sqrt{2}} \left( \frac{\partial}{\partial z_n} +
 z_n^* \right)~~~~~&;&~~~~~
\beta_n^\dagger = \frac{-1}{\sqrt{2}} \left( \frac{\partial}{\partial z_n^*} -
 z_n \right)\nonumber
\eeqa

\noindent
leading to the normal ordered (as opposed to symmeterized) Hamiltonian

\beq
H = \hbar \omega\left[ N\alpha_0^\dagger \alpha_0  + \sum_{n=1}^N
\alpha_n^\dagger \alpha_n \right].
\eeq

\noindent
Energy eigenstates can then be obtained in the usual way.  One finds the ground
state wavefunction by requiring it be annihilated by all annihilation
operators

\beqa
0 &=& \alpha_0 |0_0, 0_1, \dots, 0_N>\\
0 &=& \alpha_0 |0_0, 0_1, \dots, 0_N>.\nonumber
\eeqa

\noindent
Labeling the degeneracy of states by ``psuedoholes", the ``highest weight"
degenerate state can also be obtained by requiring it be annihilated by all
``psuedohole" annihilation operators

\beqa
0 &=& \beta_0 |0_0, 0_1, \dots, 0_N;0_0, 0_1, \dots, 0_N >\\
0 &=& \beta_0 |0_0, 0_1, \dots, 0_N;0_0, 0_1, \dots, 0_N>.\nonumber
\eeqa

\noindent
Appying creation operators then leads to the  complex wavefunction

\bdma
\lefteqn{\Psi_{\nu_1,\nu_2,\dots,\nu_N;
\mu_1,\mu_2,\dots,\mu_N}\left(z_0,z_1,\dots,z_N;
z_0^*,z_1^*,\dots,z_N^*\right)} \\
 &\equiv& \left< z_0,z_1,\dots,z_N;
z_0^*,z_1^*,\dots,z_N^*|\nu_1,\nu_2,\dots,\nu_N; \mu_1,\mu_2,\dots,\mu_N
\right>\\
&=& f_{\nu_1,\nu_2,\dots,\nu_N;
\mu_1,\mu_2,\dots,\mu_N}\left(z_0,z_1,\dots,z_N;
z_0^*,z_1^*,\dots,z_N^*\right) e^{-\frac{|z_0|^2}{N} - \sum_{n=1}^N
|z_n|^2}
\edma

\noindent
where $\nu_n$ and $\mu_n$ label the energy levels and ``psuedoholes",
respectively

\beqa
\alpha_0^\dagger \alpha_0 |\nu_0> = \nu_0 |\nu_0> ~~~&;&~~~
\alpha_n^\dagger \alpha_n |\nu_n> = \nu_n |\nu_n>\\
\beta_0^\dagger \beta_0 |\mu_0> = \mu_0 |\mu_0> ~~~&;&~~~
\beta_n^\dagger \beta_n |\mu_n> = \mu_n |\mu_n>\nonumber
\eeqa

\noindent
The functional form of the pre-exponential factor $f$ at this point in
undetermined; however it should obviously exhibit the proper statistics of
the $N$ particles.  Such statistical phase factors might be incorporated
through a modification of the oscillator commutation relations (as in the
generalization of the bosonic string to the fermionic and superstring); however
we content ourselves here with the bosonic formalism in which the bosonic
vertex operators allow for anyonic statistics.

\vspace{24pt}
\noindent
{\bf V.~~Commutation Relations and the OPE}
\vspace{12pt}

In the procedure for first quantization, the usual methods of functions and
derivatives were used as a realization of the commutator algebra for
quantum operators.  In the field-theoretic formulation of the many-body
system, it becomes convenient to realize the quantum algebra through a
mode expansion.  In the language of CFT the mode expansion is know as the
operator product expansion (OPE).  Another technique of CFT which we shall
find useful to implement is that of the construction of primary fields or
vertex operators.  In particular, winding state vertex operators have the
important property that they create coherent states for the COM motion of the
``quasiparticle".

To motivate the field-theoretic formulation, we define a quasiparticle
field $\pi (\zeta)$ in terms of an $N$-particle mode expansion

\beq
\pi(\zeta) \sim \sum_{n=0}^N \alpha_n \zeta^{n-1}
\eeq

\noindent
where $\{\alpha_n\}_{n=0}^N$ are some $N+1$ independent modes of the
quasiparticle which may be taken to be the creation-annihilation operators
of the COM conjugate momenta $(\alpha_0 \sim \pi_0^1 + i \pi_0^2)$ and the
conjugate momenta for each constituent particle of the quasiparticle
$(\alpha_n \sim \pi_n^1 + i \pi_n^2)$.  Here $\zeta$ is a complex coordinate.
In a holomorphic representation where Fourier modes are orthogonal, the
coefficient $\zeta^n$ may be loosely interpreted as the relative coordinate of
the $n$th particle from the COM of the quasiparticle.  It is related to the
fluctuation of the magnetic field ($\vec{p} \cdot \vec{A}$) from the average
magnetic field strength around the COM.  To arrive at this interpretation, we
consider our effective Hamiltonian to be

\beqa
H &=& \frac{1}{2\mu} \left(\vec{\pi}_0^2 + \sum_{n=1}^N \vec{\pi}_n^2
\right)\\
 &=& \hbar \omega \left( N \alpha_0^* \alpha_0 + \sum_{n=1}^N
{\alpha_n'}^* \alpha_n' \right)\nonumber
\eeqa

\noindent
where

\beqa
{\vec{\pi}_n}' = \vec{\pi}_n - \vec{\pi}_0~~~~&;&~~~~{\vec{\pi}_n}' =
{\vec{p}_n}' + \frac{e}{c} \vec{A}' \left(\vec{x}_n\right)\\
{\vec{p}_n}' = \vec{p}_n - \vec{p}_0~~~~&;&~~~~\vec{A}'
\left(\vec{x}_n\right) \equiv \vec{A} \left(\vec{x}_n\right) - \vec{A}
\left(\vec{x}_0\right)\nonumber
\eeqa
\beqa
\alpha_n' &\equiv& \frac{1}{\sqrt{2} \hbar} \left( \rho_n'^* + \frac{e\ell}{c}
A'(z_n) \right)
= \frac{1}{\sqrt{2}} \left( \frac{\rho_n'^*}{\hbar} + \frac{2\pi
\ell}{\Phi_0} A'(z_n) \right)\\
{\alpha_n'}^* &\equiv& \frac{1}{\sqrt{2} \hbar} \left( \rho_n' +
\frac{e\ell}{c} A'^*(z_n) \right)
 = \frac{1}{\sqrt{2}} \left( \frac{\rho_n'}{\hbar} + \frac{2\pi \ell}{\Phi_0}
A'^*(z_n) \right).\nonumber
\eeqa

\noindent
 From the correspondence principle, we then quantize by defining\footnote{This
quantization is analagous to that of the background field method of QFT.  Here
the dynamical degrees of freedom are those of the fluctuations of the
background COM.}

\beqa
\rho_n' \equiv -i\hbar \frac{\partial}{\partial z_n'}~~~~~&;&~~~~~
\rho_n'^* \equiv -i\hbar \frac{\partial}{\partial z_n'^*}.
\eeqa

\noindent
We may alternatively represent these quantum operators using the
techniques of complex variables and the residue theorem for derivatives

\beqa
2\pi i f(\zeta_0) &=& \oint \frac{f(z)}{\zeta - \zeta_0} d\zeta\\
2\pi i f^{(n)}(\zeta_0) &=& \oint \frac{f(z)}{(\zeta - \zeta_0)^{n+1}}
d\zeta.\nonumber
\eeqa

\noindent
Thus, if we represent

\beqa
\rho_n' \sim \frac{1}{\zeta^{n+1}}~~~~~&;&~~~~~ z_n' \sim \zeta^n
\eeqa

\noindent
then the canonical commutation relations can be represented by a contour
integral

\beqa
[z'_n, \rho'_m] &\equiv& \frac{1}{2\pi i} \oint \zeta^n \rho'_m(\zeta)
d\zeta\\
 &\sim& \frac{1}{2\pi i} \oint \zeta^n \zeta^{-m-1} d\zeta \sim
\delta_{nm}.\nonumber
\eeqa

\vspace{24pt}
\noindent
{\bf VI.~~Collective Coordinates for Quasiparticle States}
\vspace{12pt}

In the previous section we showed how the $N$-particle Hamiltonian could
be expressed nicely in terms of a set of COM oscillators and fluctuations from
the COM.  From this point of view the above Hamiltonian is an $N$-particle
Hamiltonian; on the other hand, we would like to reinterpret this
Hamiltonian as that of a single quasiparticle.  For this purpose it is
convenient to define a set of collective coordinates for excitations of the
quasiparticle (as opposed to the individual excitations of the individual
particles).

Assuming the current-current interaction gives rise to coherent
$N$-particle states, we define collective coordinates

\beqa
\varphi_L (\zeta) &\equiv& \varphi_{0_L} + i \sqrt{N} \alpha_0 \ln \zeta +
\sum_{n=1}^N \frac{1}{n} \left( \alpha_n(\tau) \zeta^n + \alpha_n^*(\tau)
\zeta^{-n} \right)\\
\pi(\zeta) &\equiv& \frac{\partial \varphi (\zeta)}{\partial \zeta}
 = i \sqrt{N} \frac{\alpha_0}{\zeta} + \sum _{n=1}^N \frac{1}{\zeta} \left(
\alpha_n\zeta^n - \alpha_n^* \zeta^{-n} \right).\labeq{current}
\eeqa

\noindent
These collective coordinates may alternatively be viewed as a field of
``spins" associated with the cyclotron motion of the charged particles in the
strong external magnetic field.  From this point of view, the annihilation -
creation operators ($\alpha_n$, $\alpha_n^\dagger$)  may be viewed as spin
operators annihilating and creating spins whose spin varies with $n$.

It is useful to define creation operators with negative index

\beq
\alpha^*_n \equiv \alpha_{-n}
\eeq

\noindent
allowing us to write our mode expansion in more compact form

\beqa
\varphi_L(\zeta) &\equiv& \varphi_{0_L} + i \sqrt{N}\alpha_0 \ln \zeta +
{\sum^{\textstyle \prime}}_{n=-N}^N \frac{1}{n}\alpha_n\zeta^n\\
\pi_L(\zeta) &\equiv& \frac{\varphi_L}{\partial \zeta} = \frac{1}{\zeta}
\left[ i\sqrt{N} \alpha_0 + {\sum^{\textstyle \prime}}_{n=-N}^N  \alpha_n
\zeta^n \right]\nonumber
\eeqa

\noindent
where $\sum^\prime$ indicates a sum over non-zero modes.

In our attempt to understand the critical behaviour of this system of charged
particles in a strong magnetic field, it is natural to define a stress-energy
tensor from this collective coordinate $\varphi_L(\zeta)$ which serves as
our order parameter

\beq
T_{\zeta \zeta} \equiv \pi_L^2 = \frac{\varphi_L}{\partial \zeta}
\frac{\varphi_L}{\partial \zeta}.
\eeq

\noindent
We may then associate a Hamiltonian density for an effective field theory
with this stress-energy tensor.  A natural way to do this is suggested by the
methods of CFT.  Assuming that the ground state of our system exhibits an
aproximate conformal invariance associated with a critical value for the
magnetic field strength $B_c$, it is then appropriate to identify our
Hamiltonian density with the second order moment of the stress-energy
tensor.\footnote{In this way, an integration over the two-dimensional
configuration space will yield the energy of the system.}  This second order
moment may be identified with the generator of dilations $L_0$ in a CFT.
The $n$-th order moments are called Virasoro generators.  They are the
generators of $(n+2)$-th order conformal transformations in such a field
theory [\rref{BPZ}] defined by

\beq
T(\zeta) \equiv \sum_{n=-\infty}^\infty \frac{L_n}{\zeta^{n+2}}.
\eeq

\noindent
Expanding our stress-energy tensor, we then have

\beqa
T(\zeta) &=& \frac{1}{\zeta^2} \left[ i \sqrt{N} \alpha_0 + \sum_{n=1}^N
\left( \alpha_0 \zeta^n - \alpha^* \zeta^{-n} \right) \right]^2\\
 &=& \frac{1}{\zeta^2} \left\{ \left. -N \alpha_0^2 + 2i \sqrt{N} \alpha_0
\sum_{n=1}^N \left(\alpha_n \zeta^n - \alpha_n^* \zeta^{-n} \right) \right.
\right.\nonumber\\
 & & \left.  + \sum_{n,m=1}^N \left[ \alpha_n \alpha_m \zeta^{n+m} +
\alpha_n^*
\alpha_m^* \zeta^{-n-m} - \left( \alpha_n^* \alpha_m \zeta^{m-n} + \alpha_n
\alpha_m^* \zeta^{n-m} \right) \right] \right\}\nonumber
\eeqa

\noindent
Thus, in the large $N$-limit, we have

\beq
L_0 \approx -N \alpha_0^2 - \sum _{n=1}^N \left[ \left( \alpha_n \alpha_{-n} +
\alpha_n^* \alpha_{-n}^* \right) + \left( \alpha_n^* \alpha_n + \alpha_n
\alpha_n^* \right) \right]
\eeq

\noindent
which, when ``normal ordered" in operator form reduces to

\beq
L_0 \approx N \alpha_0^* \alpha_0 + {\sum^{\textstyle \prime}}_{n=-N}^N
\alpha_n^* \alpha_n.
\eeq

\noindent
We thus define our Hamiltonian (density) as

\beqa
H &=& N\alpha_0^* \alpha_0 + \eta\nonumber\\
\eta &\equiv& {\sum^{\textstyle \prime}}_{n=-N}^N \alpha_n^* \alpha_n\\
H |{\rm phys}> &=& \left(N\alpha_0^* \alpha_0 + \eta \right) |{\rm
phys}>\nonumber
\eeqa

\noindent
where energy eigenstate may be labeled by the COM momentum $\alpha_0$
and $|{\rm phys}>$ are ``physical" states whose characteristics and
significance are yet to
be determined.  In particular we expect BRST transformations to
play an important role here.  BRST transformations ($Q$) are related to
the reparameterization invariance associated with the Virasoro algebra.  In
this formalism, physical states are related to ``exact" states whose
$Q^2$ eigenvalue vanishes.  Physical ``null" states are related to closed, but
not exact physical states.  We expect these null states to be related to states
which are just ``gauge equivalent" to the exact states.  Assuming the
techniques of CFT to be applicable, we thus expect the approximate
conformal symmetry  to tell us something about the transformation (and
thus internal symmetry) properties of the wavefunctions associated with the
various excitations of the system.

\vspace{24pt}
\noindent
{\bf VII.~~Collective Coordinates for ``Quasipsuedoholes"}
\vspace{12pt}

We find it useful to define collective coordinates for ``quasipsuedoholes" in
order to understand the symmetries associated with the Landau degeneracy.
 Recall that for a single (two-dimensional) particle in an external magnetic
field, there exists a conserved quantity ($\tilde{\pi}$) associated with the
bariocentric coordinate for the cyclotronic motion of the particle.  Since this
conserved quantity can be represented by the charge conjugation of the
canonical momentum, we have called this a ``psuedohole".\footnote{The
conserved charges have different interpretations in different gauges.  Our
description in terms of a ``psuedohole" comes from the symmetric gauge
choice.  This obviously generalizes to other gauges.}

Following the quasiparticle case, we define

\beqa
\tilde{\pi} (\zeta^*) &\equiv& \frac{\partial \varphi_R (\zeta^*)}{\partial
\zeta^*}\nonumber\\
\varphi_R (\zeta^*) &\equiv& \varphi_{0_R} + i \sqrt{N} \beta_0 \ln \zeta^* +
{\sum^{\textstyle \prime}}_{n=-N}^N \frac{1}{n} \beta_n {\zeta^*}^n\\
\tilde{T}_{\zeta^* \zeta^*} &=& \tilde{\pi}^2\nonumber\\
\tilde{H} &=&\tilde{N} \beta_0^* \beta_0 + {\sum^{\textstyle \prime}}_{n=-N}^N
\beta_n^*\beta_n
\sim \tilde{L}_0\nonumber
\eeqa

\noindent
where $\tilde{L}_0$ is the second order moment of the stress energy tensor
$\tilde{T}_{\zeta^* \zeta^*}$.
This quasipsuedohole Hamiltonian may be considered to describe the
``dynamics" of the bariocentric coordinates associated with the position of a
magnetic flux quantum.  Since the complex coordinates label the positions of
the particles (as in the usual formulations of CFT's),  the generator of
rotations is easily shown to be

\beq
L_3 \equiv \hbar \left( L_0 - \tilde{L}_0 \right).
\eeq

\vspace{24pt}
\noindent
{\bf VIII.~~Conclusion}
\vspace{12pt}

We have shown how the addition of a ``current-current" interaction between
particles leads to a diagonalization of the $N$-particle Hamiltonian.
Interpreting this interaction term as that associated with fluctuations from
the COM leads to a field theoretical Hamiltonian resembling that of a 2+1
dimensional CFT  in the large $N$ limit.  A similar field theoretical
Hamiltonian may also be constructed for the states associated with the
``psuedohole" operators.  It should thus be possible to construct a 2+1
dimensional field theory incorporating the dynamics of both particle and
``psuedohole" fluctuations.  Such a field theory could be useful in developing
a nonlinear framework from which it might be possible to understand the
connections between different ground state vacuua associated with the
different phases of the QHE.

\vspace{48pt}
\noindent
{\bf Acknowledgements}
\vspace{12pt}

It is a pleasure to thank S. Thomas, J. Gaite and Q. Niu for helpful
discussions.  This work was supported by DOE grant DOE-287-CPP-25.

\newpage
\begin{center}
{\large {\bf References}}
\end{center}

\begin{enumerate}




\item Laughlin R.B., Phys. Lett. {\bf 50} (1983) 1395.\labref{L3}


\item Fubini S., {\em Vertex Operators and the Quantum Hall Effect},
CERN-TH. 5922/90.\\
Fubini S. and L\"{u}tken C.A., {\em Vertex Operators in the Fractional
Quantum Hall Effect}, CERN-TH. 5960/90.\labref{F2}

\item Stone M., {\em Vertex Operators in the Quantum Hall Effect}, IL-
TH-90-\# 32.\labref{S1}\\
Balatsky A. and Stone M., {\em Vertex Operators and Spinon Edge
Excitations in the Spin-Singlet Quantum Hall Effect}, IL-TH-90-\#35.\\
Dunne G.V., Lerda A. and Trugenberger C.A., {\em Landau Levels and
Vertex Operators for Anyons}, CTP \# 1938 (and LANL 91-535).


\item Wen X.G., {\em Gapless Boundary Excitations in the Quantum Hall States
and in the Chiral Spin States}, NSF-ITP-89-157 (unpublished).\labref{Wen1}

\item Nagao G., {\em ``Psuedoholes", Gauge Constraints and Critical Phenomena
in the Quantum Hall Effect}, unpublished. \labref{N1}

\item Belavin A.A., Polyakov A.M. and Zamolodchikov A.B., {\em Infinite
Conformal Symmetry In Two-Dimensional Quantum Field Theory}, Nucl.
Phys. {\bf B 241} (1984) 333.  \labref{BPZ}

\end{enumerate}

\end{document}